\documentstyle[aps,epsfig,prl,preprint]{revtex}

\begin{document}
\draft
%@twocolumnfalse\endcsname
\title{Observation of Spin Freezing in  CaV$_4$O$_9$ and CaV$_2$O$_5$ by $\mu$SR}
\author{G.~M.~Luke, Y.~Fudamoto, K.~M.~Kojima, M.~Larkin, J.~Merrin, B.~Nachumi, S.~Sinawi
 and Y.~J.~Uemura}
\address{Dept. of Physics, Columbia University, New York, NY 10027, U.~S.~A.}
\author{M.~J.~P.~Gingras}
\address{Dept. of Physics, University of Waterloo, Waterloo, Ont. N2L-3G1, 
Canada.}
\author{M.~Sato and S.~Taniguchi}
\address{Dept. of Physics, Nagoya University, Furo-cho, Chikusa-ku,
Nagoya 464-1, Japan and CREST, Japan Science and Technology Corporation (JST).}
\author{M.~Isobe and Y.~Ueda}
\address{ISSP, University of Tokyo, Roppongi 7-22-1, Minato-ku, Tokyo 106, Japan.}
\date{\today}
\maketitle
\begin{abstract}
We have performed muon spin relaxation and susceptibility
 measurements of  CaV$_4$O$_9$ 
and CaV$_2$O$_5$, systems in which  spin gap-like behavior has been found.
  We observe spin freezing in the two
compounds below 12~K and 50~K, respectively.  Our results indicate
the possibility
that a substantial fraction of the vanadium magnetic moment is not in
a spin singlet state at low temperatures in either material.
\end{abstract}
\pacs{75.50.Lk, 76.75.+i}
%\vskip2pc]
%\narrowtext
The properties of low dimensional and/or frustrated magnetic
 materials have attracted considerable
attention due to the appearance of various  
 non-N\'eel ground states.  Systems with singlet ground states and spin gaps
include the one-dimensional Haldane (integer spin chain,
 {\em eg.}Y$_2$BaNiO$_5$) and
spin-Peierls (spin 1/2 chain, {\em eg.} CuGeO$_3$) states.
  Spin ladders with even numbers of legs (such as SrCu$_2$O$_3$)
can  also have spin 
singlet ground states. 
Recently, a possible new type of singlet ground state, the plaquette resonating
valence bond (PRVB) state, where the singlet is comprised of four spins in a
two dimensional arrangement, has received considerable theoretical 
attention\cite{nk95a,ks96,ku96a,mt96a,tm96,oas96,mpg96,ss96a,srw96}
since the proposal by Taniguchi {\em et al.}\cite{st95a} that
CaV$_4$O$_9$ possesses such a ground state.

AV$_n$O$_{2n+1}$ ($n=2,3,4$, A=Ca, Na, Sr)  is  a series of related
compounds in which stacked layers of edge sharing VO$_5$
pyramids form an essentially square lattice in 
which $1/(n+1)$ of the spin-1/2 V lattice points are periodically
depleted.\cite{jcb76}
An  extremely interesting feature of these vanadates  is that they
 offer the possiblity of studying several different magnetic
ground states in a family of related compounds.
In CaV$_3$O$_7$ ($n=3$)
all the vanadium ions are in the spin 1/2 V$^{4+}$ state and order 
antiferromagnetically\cite{gl93a,hh96a}.
In CaV$_2$O$_5$ the V$^{4+}$ ions are arranged in the
two leg ladder structure
while spins in adjacent ladders are coupled via
 a zig-zag chain structure.
A spin gap has been inferred from NMR and susceptibility in CaV$_2$O$_5$,
although its microscopic  origin 
isn't clear, as it is unknown which exchange interaction
(ladder or zig-zag) is dominant\cite{hi96}.
The structure of NaV$_2$O$_5$ is similar to that of CaV$_2$O$_5$, 
except that
half of the V ions are V$^{4+}$ (spin 1/2) while the
other half are V$^{5+}$ (spin 0), resulting in a 1-d chain structure;
a spin-Peierls transition at 34~K (the highest yet observed) has
recently been identified\cite{mi96}.
  Susceptibility\cite{st95a}, NMR--1/T$_1$\cite{st95a,to97,kk96}
and inelastic neutron scattering\cite{kk96} measurements
of CaV$_4$O$_9$ ($n=4$) have indicated the presence of a
spin gap   which has generated considerable theoretical
interest as this would be the first two dimensional system
 with a spin singlet ground state.

The first proposed mechanism for the singlet state in CaV$_4$O$_9$ was
the  plaquete resonating valence bond (PRVB) state, which consists
of local singlets from the four spins on a nearest or next-nearest
neighbour plaquette.
Depending on the strength of the various exchange interactions, 
different ground states, including N\'{e}el order, dimerization and
the PRVB state have been found theoretically;
 several authors have argued that frustration effects 
select the PRVB state.  
Orbital ordering has been suggested as a spin gap mechanism in
both  CaV$_2$O$_5$
and CaV$_4$O$_9$\cite{sm97}.  However,
in the case of CaV$_4$O$_9$, this suggestion appears to be inconsistent
with neutron\cite{kk96}and  susceptibility\cite{mpg97} measurements.

Ceramic  specimens of  CaV$_4$O$_9$ and CaV$_2$O$_5$ were grown
at Nagoya University and ISSP respectively,  as described
 previously\cite{st95a,hi96}. 
Field-cooled susceptibility measurements
are shown in Fig.~1. For
 CaV$_4$O$_9$,
$\chi$ shows a broad maximum around 100K, then drops rapidly with
 decreasing temperature,
while for CaV$_2$O$_5$, $\chi$ peaks above room temperature then
 becomes essentially temperature 
independent below about 80~K.
   A weak Curie term
in both samples 
at low temperatures corresponds to an impurity level of $\sim0.1$\%
 V$^{4+}$ free moments.

We performed detailed dc-susceptibility measurements to search for
irreversibility (a signature of spin glass order)
in both samples.  We first cooled
in zero field to 5~K, applied a magnetic field, then took measurements
in increasing temperature up to room temperature, and then upon cooling to
low temperatures.  No irreversibility was observed in fields of
100~G, 1000~G or 1~T for CaV$_4$O$_9$.
In the case of CaV$_2$O$_5$,
we found that irreversibility set in at about 50~K, as shown in Fig.~1.
  We performed this measurement sequence in a series of fields,
finding that the irreversibility was greatest in an applied field of 
roughly 1~kG.  The degree of irreversibility depended on the field as well
as the time delay at each point, with a maximum of about 15~\% for the
fractional change in the moment.  Irreversibility of greater than 2~\%
was apparent for fields between 500~G and 1~T.

Muon spin relaxation ($\mu$SR) is an extremely sensitive magnetic probe.
  Internal
magnetic fields of several gauss (corresponding to typical static
moments less than  $0.1\;\mu_B$) are readily detected.
As a real-space probe, $\mu$SR  can sensitively
determine ordered  volume fractions  and is especially powerful
for the study of disordered magnetic materials such as spin glasses.  
 The specimens were mounted
on a pure silver sample holder in either a helium gas flow cryostat or dilution
refrigerator at the M13 and M15
surface muon channels at TRIUMF.   Spectra were measured in zero field
and longitudinal field geometries.

%===========================
%	CaV_4O_9
%============================

 We had anticipated confirming the non-magnetic ground state in
CaV$_4$O$_9$ and CaV$_2$O$_5$ by observing only weak, temperature-independent
 muon spin relaxation from static nuclear dipole moments and possibly some 
weak dynamic relaxation due to free moments associated with local defects
in the singlet ground state as in our previous studies of other
spin singlet systems\cite{kk95a,ot95,kk95b}.
  Instead, below 100~K in CaV$_4$O$_9$ and
150~K in CaV$_2$O$_5$,
we observed increasingly strong dynamic relaxation in both zero field and in
an applied longitudinal field as shown in  
Fig.~2a for  CaV$_4$O$_9$; spectra
 for CaV$_2$O$_5$ are qualitatively similar.  
In both materials the
 relaxation rate increased with decreasing temperature, indicating
the slowing down of fluctuating paramagnetic moments.
Above 15~K in CaV$_4$O$_9$ and 80~K in CaV$_2$O$_5$,
 the relaxation was exponential while it
 decayed with a two component
form at lower temperatures. 

Fig.~3 shows
the relaxation rates for the two vanadates
 in both low field (zero field or 100~G)  and
high field (3000~G in CaV$_4$O$_9$ and 1000~G in CaV$_2$O$_5$).
Filled symbols are fits to a single exponential relaxation rate (used for
all fields at higher temperatures and low fields at low temperature) while
open symbols are the relaxation rate of the longer-lived tail component
(which reflects the effects of fluctuating fields) in high fields at lower
temperatures.   The relaxation rate of the longer-lived component reaches
a maximum value at 50~K in CaV$_2$O$_5$ and at 12~K in CaV$_4$O$_9$, then
decreases with decreasing temperature. 
The weak field dependence at high temperature is characteristic
 of a paramagnetic state, while the two component form,
where the application of a longitudinal field decouples the fast 
relaxing signal,
increasing the amplitude of the slowly relaxing component (as shown in 
Fig.~2b),
 indicates that the 
source of the relaxation at low temperatures in both materials
is a quasi-static distribution of internal magnetic
fields, such as is found in a spin glass.  All of this behavior
seen in $\mu$SR 
indicates that CaV$_4$O$_9$ and CaV$_2$O$_5$ undergo spin freezing
below 12~K and 50~K, respectively.

  In both materials, the amplitude
of the relaxing signal indicates that essentially the entire volume fraction 
($>95$~\%) is involved in the spin freezing.
 We can estimate the magnitude of the
characteristic  quasi-static internal fields at low temperatures
from  the zero/low field relaxation
rates, obtaining $20~\mu{\rm s}^{-1}/\gamma_\mu\sim250$~G for CaV$_4$O$_9$ and
65~$\mu{\rm s}^{-1}/\gamma_\mu\sim760$~G for CaV$_2$O$_5$. 

In the paramagnetic regime, the
  muon spin relaxation rate in an applied longitudinal field
$B_L=\frac{\omega_L}{\gamma_\mu}$  due to electronic moments 
fluctuating with an inverse correlation time $\nu$
is given by
\begin{equation}
1/T_1 = \frac{\gamma_\mu^2B_{{\rm inst}}^2\nu}{\nu^2+\omega_L^2}, 
\;\;\;\;\;\;\nu\gg\gamma_\mu B_{{\rm inst}}
\label{tone}
\end{equation}
 where  $B_{{\rm inst}}$ is the average size of the instantaneous
local field, $\gamma_\mu=85.1$~kHz/G is the muon gyromagnetic ratio.
  Measurement
of $1/T_1$  for several fields at the same temperature allows the
simultaneous determination of $B_{{\rm inst}}$ and $\nu$.

We attempted to use Eqn.~\ref{tone} to calculate the size and fluctuation
rate of the local fields in CaV$_2$O$_5$ above 50~K.
  However, we found that the
relaxation rate actually increased with increasing field around 1~kG,
which is inconsistent with the expected behavior for paramagnetic moments.
This effect indicates that the application of the field actually
enhanced the slowing down of the fluctuations. 
If the spin freezing depends on  the frustration of antiferromagnetic
interactions by competing ferromagnetic interactions, then the application
of a field could enhance this and promote spin freezing.

We measured the $1/T_1$ relaxation rate in CaV$_4$O$_9$
for a range of  applied fields between 100~G and 3500~G,
 at five temperatures between 18~K and 30~K (shown in Fig.~4).
Fitting $1/T_1$  to Eqn.~1 for each temperature independently
we obtained $\gamma_\mu B_{{\rm inst}}=21\pm0.3\;\mu{\rm s}^{-1}$ for
 each. This value of $B_{{\rm inst}}$ also agreed with 
 the static local field seen at low temperature in this material,
indicating that there is no change in the size of the moments responsible
for the spin relaxation between 2~K and 30~K.  Using this value
of $B_{{\rm inst}}$,
 we obtain the temperature dependence of the
fluctuation rate $\nu$, shown in the inset of Fig.~4.

%===

The fluctuation rate $\nu$ increases quickly with increasing temperature
at low temperatures, then eventually begins to saturate at higher 
temperatures.  As shown in the inset of Fig.~4, we
 can fit the entire temperature dependence with the 
phenomenological form:
\begin{equation}
\nu = \nu_\infty\frac{1}{e^{\Delta/kT}+1}
\label{nu}
\end{equation}
where we find $\nu_\infty=(3.0\pm0.3)\times10^{10}\;{\rm s}^{-1}$ 
and $\Delta=114\pm3$~K.
This form would be reasonable if the fluctuation rate were determined by the 
thermal population of some excited state, with an energy gap $\Delta$ above
the ground state.  The gap value that we obtain is in good agreement with
the values seen with susceptibility, NMR and inelastic neutron scattering which
suggests that we are 
observing spin dynamics which are coupled to those seen in those
techniques.  However, the overall behaviour we see is different from
these other techniques: for example, the relaxation rate in NMR
decreases with  decreasing temperature, here, it increases.
 
The temperature dependence of $\nu$ given by Eqn.~\ref{nu}
 would
be expected if there were two different spin systems, one which relaxed
the other.  As the subsystem causing the cross-relaxation were gapped, then
the remaining subsystem would no longer be relaxed and could
 undergo slowing down and eventual spin freezing.  Additional evidence for
the presence of two different spin subsystems is the slow fluctuation rate
$\nu_\infty=3\times10^{10}\;{\rm s}^{-1}$, much lower than estimates
of the exchange frequencies  ($\sim3\times10^{12}\;{\rm s}^{-1}$) in these
materials.

%  
%=====

%=========================
%	now CaV2O5
%=========================

%We first note that we are able to measure the field fluctuation rate $\nu$
% over a wide range of temperature with $\mu$SR.  In the case of critical
%%slowing down above an antiferromagnetic transition (such as in CaV$_3$O$_7$),
%slow fluctuations which are able to produce measurable T$_1$ 
%relaxation within the $\mu$SR time window are seen only for
%less than  a few Kelvin
%above the ordering temperature.  The observation of slow fluctuations over
%such a wide temperature range as we see here is generally
% characteristic of a highly
%frustrated spin system.

Our observations of slowing down of paramagnetic spin fluctuations over a
wide temperature range and spin freezing below 12 ~K in CaV$_4$O$_9$
and 50~K in CaV$_2$O$_5$ are at first difficult to reconcile  with a
 picture  of a simple spin singlet ground state in the two systems.
The size of the characteristic local fields in the two compounds (250~G 
and 760~G, respectively) are comparable in magnitude to the uniform fields 
seen in the antiferromagnetic
state of CaV$_3$O$_7$ (600~G and 1.5~kG for the two muon sites in that
 material).  Since each V$^{4+}$ spin in CaV$_3$O$_7$
has an ordered moment of roughly 0.4~$\mu_B$,\cite{hh96a}
 we see that a sizable fraction
of the V moments in CaV$_4$O$_9$ and CaV$_2$O$_5$ must have 
frozen moments on the order of 0.1~$\mu_B$. Thus, we can exclude
 the possibility of spin freezing of   dilute
 impurities (such as seen in susceptibility)
causing the observed phenomena.  One might imagine that the implanted muons
could act as impurities and induce spins around the muon.
  However, such a case has 
not 
been seen in other singlet systems, including Haldane, spin-Peierls
 or  copper based
spin ladders. If spins are actually induced by the muons it
would indicate an extremely sensitive nature of the ground state of the
host spin systems.  The observation of irreversibility 
 in the
susceptibility  of CaV$_2$O$_5$ at the same temperature
as the spin freezing in $\mu$SR
 argues against the possibility of
a muon-induced effect in that compound.

 A similar situation
 arose in studies of the two leg ladder system
LaCuO$_{2.5}$, where spin gap behavior  (with $\Delta=474$~K)
 was seen in susceptibility 
measurements\cite{zh95}.
  However, NMR and  $\mu$SR measurements found magnetic ordering
below T$_N=120$~K\cite{rk96}.  Subsequent theoretical calculations
\cite{mt97}  showed that  near 
the quantum critical point of a transition to a spin liquid 
 $\chi(T)=\chi_0+aT^2$, where $\chi_0$ can be
small.  As a result, one  observes 
a large decrease in $\chi$ which mimics a spin gap, yet still
finds a substantial ordered moment.
 Calculations  of
the unfrustrated  CaV$_4$O$_9$ lattice find N\'{e}el order\cite{mt96a}
for a wide range of couplings yet the calculated $\chi(T)$
 would be difficult to distinguish from
a true gap due to
 experimental difficulties in determining $\chi_{{\rm spin}}=0$
in the presence of temperature-independent contributions (Van-Vleck,
core diamagnetism etc.) and Curie-like impurity contributions.

Inelastic neutron scattering measurements of CaV$_4$O$_9$
 have exhibited clear gap-like behavior\cite{kk96}.
However, those results do not exclude the existence of the frozen
moment seen in the present study;  the
slow fluctuations  and freezing 
we observe would be at too low energy to have
 been detected
in the neutron measurements. 
Spin freezing should be apparent in NMR measurements of both systems.
In fact, $^{51}V$ NMR measurements of CaV$_4$O$_9$ deviate from their
activated form around 15~K, although this was attributed to impurities.
Our results indicate this effect may be intrinsic.  A potential 
complication in comparing $\mu$SR and NMR measurements is 
  that the NMR experiments were performed in high
magnetic field (7T) which may affect the low temperature properties.
Zero field NQR measurements could provide valuable additional information.   

Our observation of spin freezing involving a substantial moment in both
CaV$_4$O$_9$ and CaV$_2$O$_5$ implies that the ground state of both systems
is more complicated than the spin singlets which have been proposed to date.
In CaV$_4$O$_9$, we see evidence for the presence of two spin systems:
one of which is gapped while the other undergoes spin freezing.
   It remains to be seen whether the different subsystems
correspond to distinct vanadium spins, or to different portions of 
each vanadium moment.

We appreciate the assistance of Syd Kreitzman, Bassam Hitti
 and the TRIUMF-$\mu$SR
User Facility.
Work at Columbia was supported by NSF-DMR-95-10453, 10454 and NEDO (Japan).

%\figure{Crystal structure of CaV$_4$O$_9$ and CaV$_2$O$_5$.}
\begin{figure}
\begin{center}
\epsfig{file=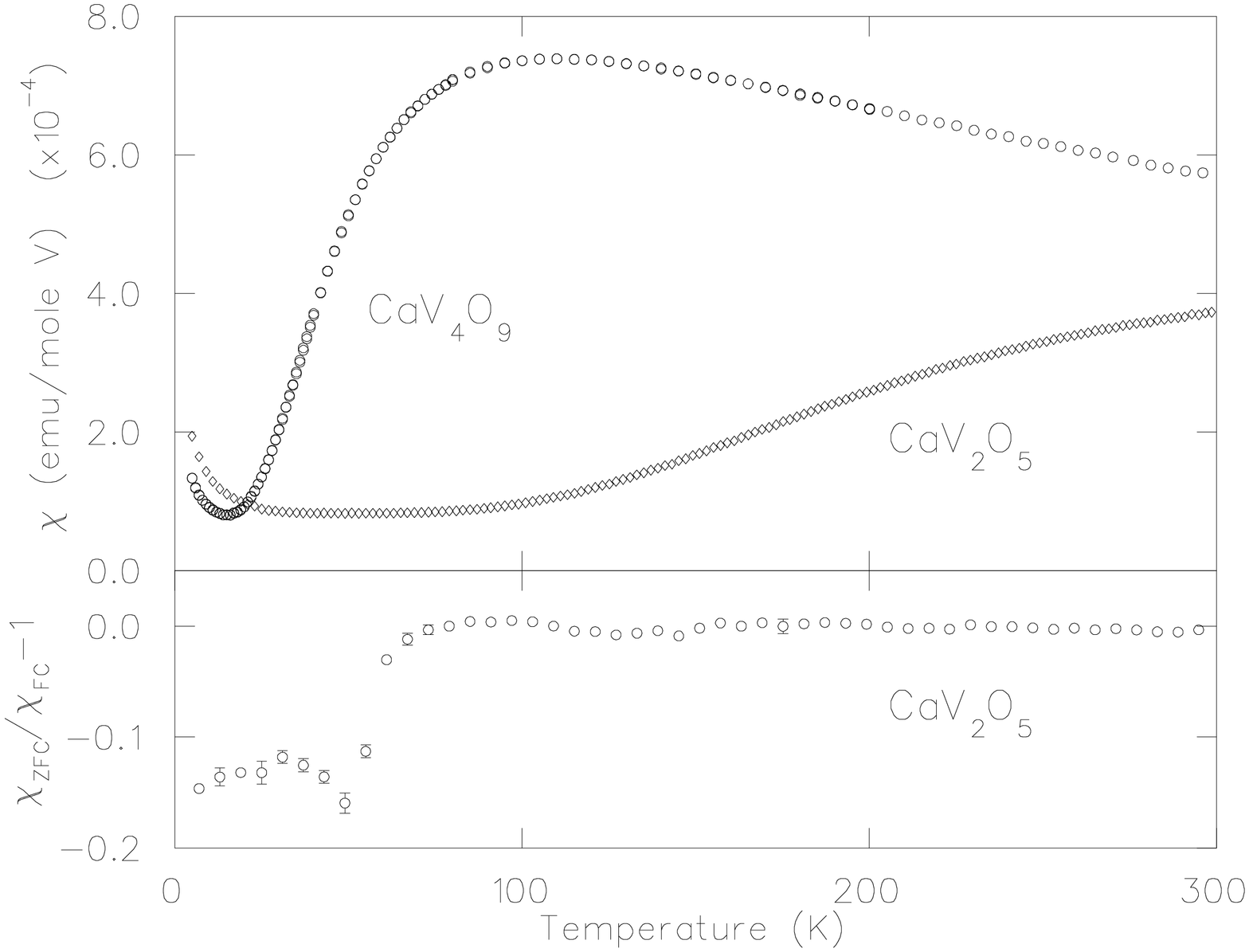,angle=90,width=\columnwidth}
\end{center}
\caption[]{(a)Magnetic susceptibilities of CaV$_4$O$_9$ and CaV$_2$O$_5$
and (b) Fractional irreversibility ($\chi_{{\rm ZFC}}/\chi_{{\rm FC}}-1$) in 
CaV$_2$O$_5$.}
\end{figure}
\begin{figure}
\begin{center}
\epsfig{file=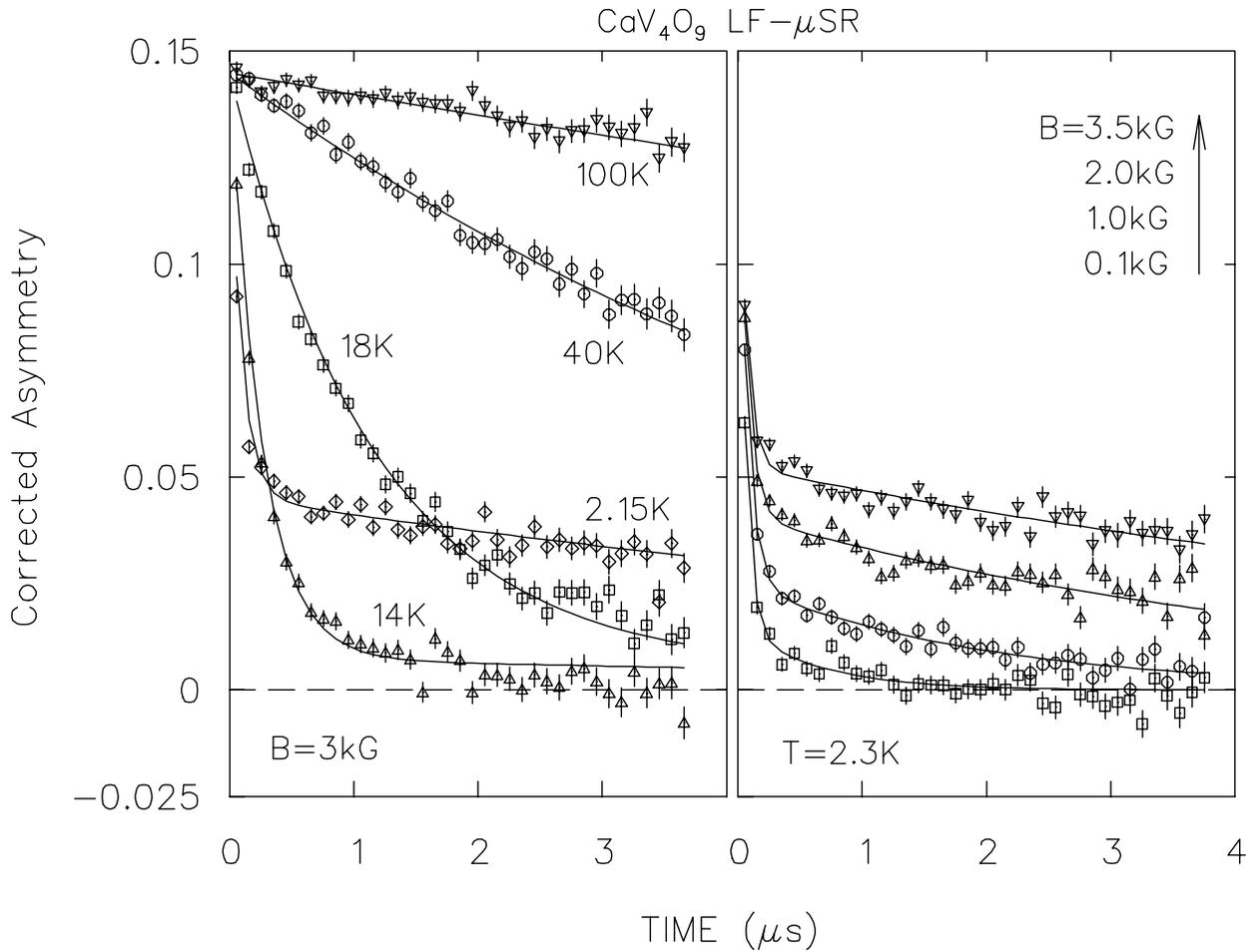,width=\columnwidth}\end{center}
\caption[]{(a)LF-$\mu$SR spectra measured in B=3000~G for CaV$_4$O$_9$ for
 T=100~K, 40~K, 18~K, 14~K and 2.15~K and (b)LF-$\mu$SR spectra measured
 at 2.3~K in
 CaV$_4$O$_9$ for B=100~G, 1000~G, 2000~G and 3500~G.}  
\end{figure}
\begin{figure}
\begin{center}
\epsfig{file=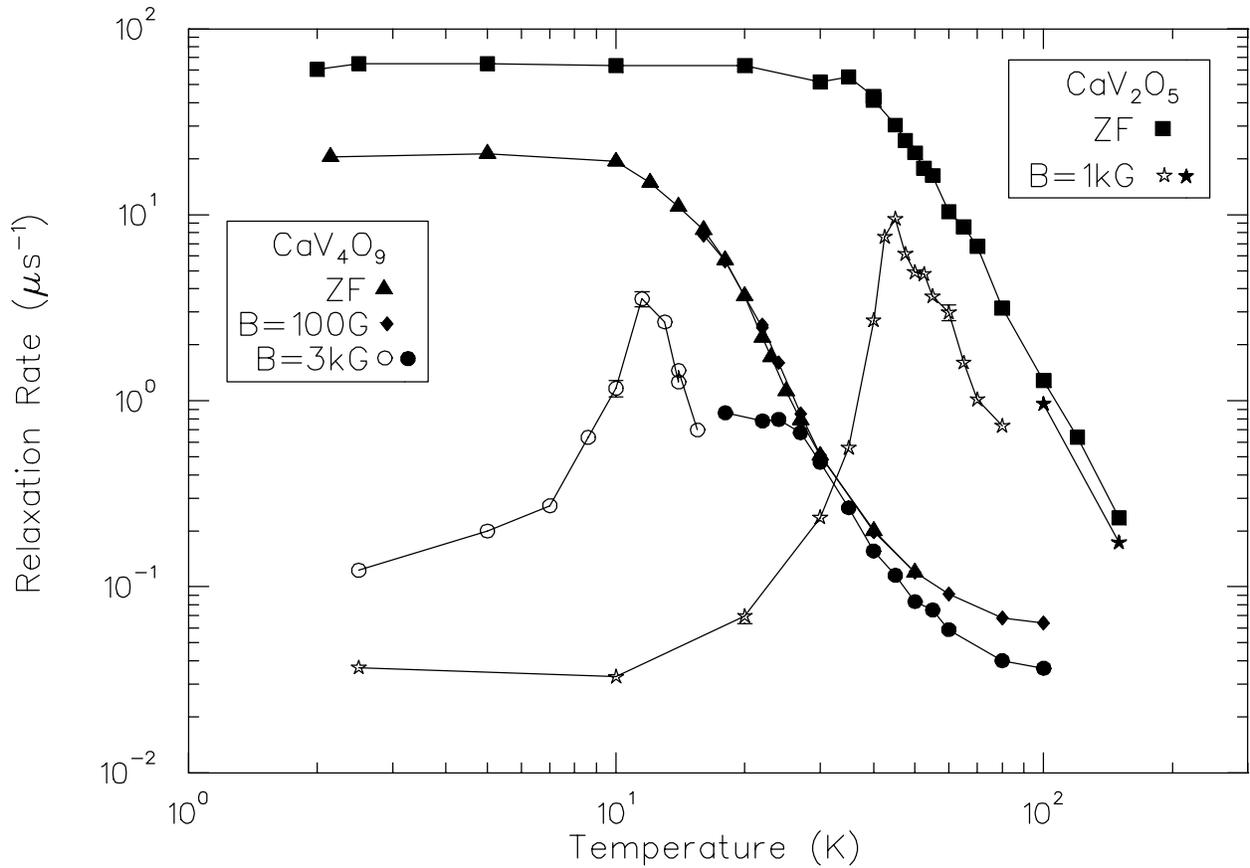,width=\columnwidth}\end{center}
\caption[]{Relaxation rate {\em vs.} temperature in: CaV$_4$O$_9$,
for zero field (triangles), 100~G (diamonds) and 3000~G (circles);
CaV$_2$O$_5$ in zero field (squares) and 1000~G (stars).
Open symbols indicate a two component relaxation function and correspond
to the long-lived  component which reflects dynamical (1/T$_1$) relaxation.}
\end{figure}
\begin{figure}
\begin{center}
\epsfig{file=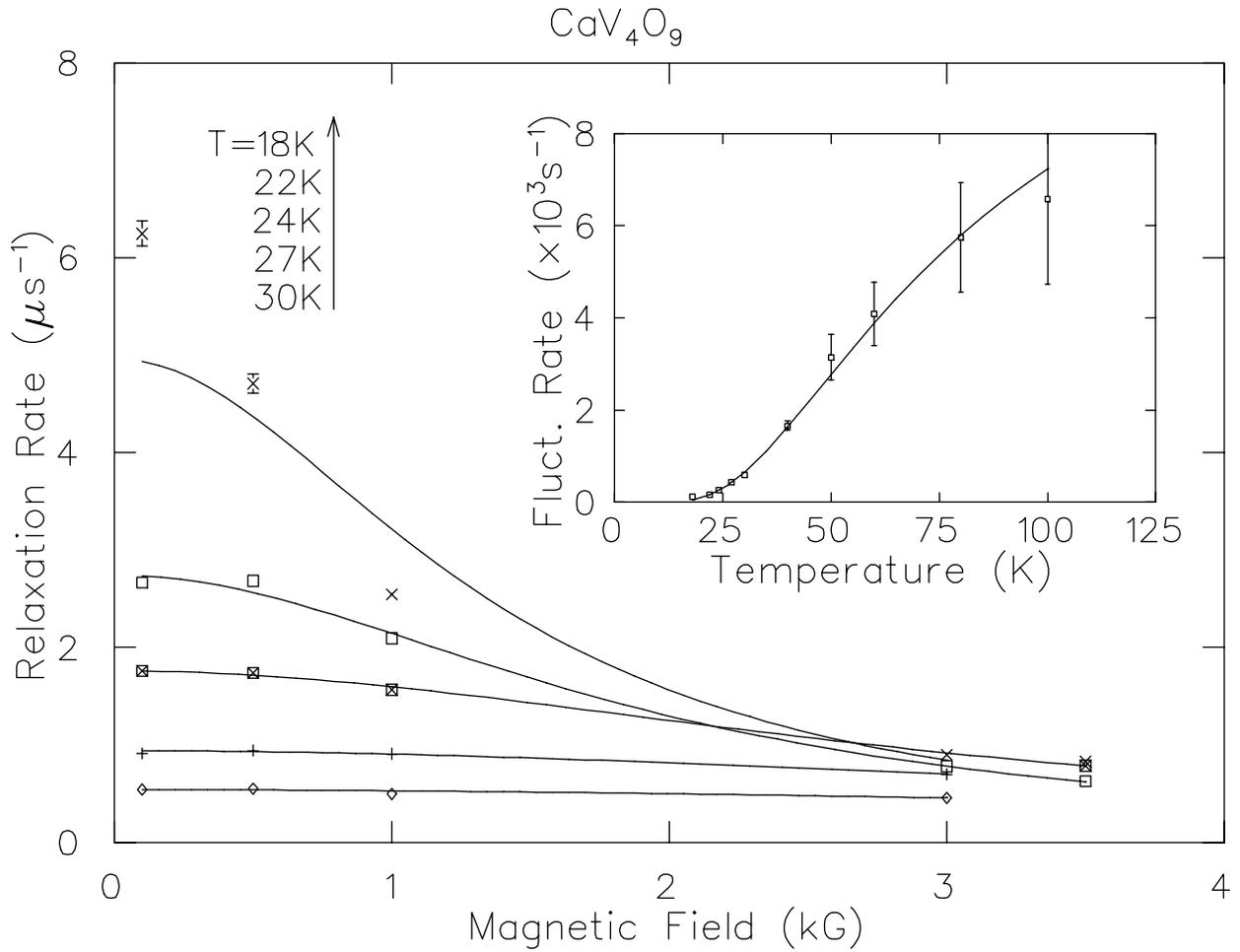,width=\columnwidth}\end{center}
\caption[]{1/T$_1$ Relaxation rate {\em vs.} field in CaV$_4$O$_9$
 for T=18~K, 22~K, 24~K, 27~K
and 30~K.  Curve is fit to Eqn.~1.  Inset: Field fluctuation rate from Eqn.~1.}
\end{figure}

\end{document}